\documentclass{comnet}

\usepackage{graphicx}

\usepackage{mathtools}

\usepackage{color}

\jno{xxxxxx}
\received{XX XX XXXX}
\revised{XX XX XXXX}
\accepted{XX XX XXXX}

\begin{document}

\title{Detecting global bridges in networks}

\shorttitle{Detecting global bridges in networks}
\shortauthorlist{Pablo Jensen et al.}

\author{ 
  \name{Pablo Jensen$^*$} 
  \address{IXXI, Institut Rhonalpin des Syst\`emes Complexes, ENS Lyon;
    Laboratoire de Physique, UMR 5672, ENS Lyon 69364 Lyon, France\email{$^*$Corresponding author: pablo.jensen@ens-lyon.fr}}
  \name{Matteo Morini}
  \address{IXXI, Institut Rhonalpin des Syst\`emes Complexes, ENS Lyon;
    LIP, INRIA, UMR 5668, ENS de Lyon 69364 Lyon, France}
  \name{M\'arton Karsai}
  \address{IXXI, Institut Rhonalpin des Syst\`emes Complexes, ENS Lyon;
    LIP, INRIA, UMR 5668, ENS de Lyon 69364 Lyon, France}
  \name{Tommaso Venturini}
  \address{M\'edialab, Sciences Po, Paris}
  \name{Alessandro Vespignani}
  \address{MoBS, Northeastern University, Boston MA 02115 USA;
    ISI Foundation, Turin 10133, Italy}
  \name{Mathieu Jacomy}
  \address{M\'edialab, Sciences Po, Paris}
  \name{Jean-Philippe Cointet}
  \address{Universit\'e Paris-Est, SenS-IFRIS}
  \name{Pierre Merckl\'e}
  \address{Centre Max Weber, UMR 5283, ENS Lyon 69364 Lyon, France}
  \name{Eric Fleury} 
  \address{IXXI, Institut Rhonalpin des Syst\`emes Complexes, ENS Lyon; 
    LIP, INRIA, UMR 5668, ENS de Lyon 69364 Lyon, France}}

\maketitle

\begin{abstract} {The identification of nodes occupying important
    positions in a network structure is crucial for the understanding
    of the associated real-world system. Usually, betweenness
    centrality is used to evaluate a node capacity to connect
    different graph regions. However, we argue here that this measure
    is not adapted for that task, as it gives equal weight to ``local'' centers
    (\emph{i.e.} nodes of high degree central to a single region) and
    to ``global'' bridges, which connect different communities. This
    distinction is important as the roles of such nodes are different
    in terms of the local and global organisation of the network
    structure. In this paper we propose a decomposition of betweenness
    centrality into two terms, one highlighting the local contributions and the other the global ones. We call the latter
    \emph{bridgeness} centrality and show that it is capable to specifically spot out global bridges. In addition, we
    introduce an effective algorithmic implementation of this measure
    and demonstrate its capability to identify global bridges
    in air transportation and scientific collaboration networks.}
  {Centrality Measures, Betweenness Centrality, Bridgeness Centrality}
\\
  JXX, JYY
\end{abstract}

\section{Introduction}

Although the history of graphs as scientific objects begins with
Euler's \cite{Euler1736} famous walk across K\"onigsberg bridges,
the notion of 'bridge' has rarely been tackled by network
theorists\footnote{We refer to the common use of the word 'bridge', 
and not to the technical meaning in graph theory as 'an edge whose
deletion increases its number of connected components'}. Among the
few articles that took bridges seriously, the most famous is probably
Mark Granovetter's paper on The Strength of Weak Ties
\cite{Granovetter1973}. Despite the huge influence of this paper, few
works have remarked that its most original insights concern precisely
the notion of 'bridge' in social networks. Granovetter suggested that
there might be a fundamental functional difference between strong and
weak ties. While strong ties promote homogeneous and isolated
communities, weak ties foster heterogeneity and crossbreeding. Or, to
use the old t\"onnesian clich\'e, strong ties generate Gemeinshaft, while
weak ties generates Gesellshaft \cite{Coser1975}. Although Granovetter
does realize that bridging is the phenomenon he is looking after, two
major difficulties prevented him from a direct operationalization of
such concept: ``We have had neither the theory nor the measurement and
sampling techniques to move sociometry from the usual small-group
level to that of larger structures'' (ibidem, p. 1360). Let's start
from ``the measurement and sampling techniques''. In order to compute
the bridging force of a given node or link, one needs to be able to
draw a sufficiently comprehensive graph of the system under
investigation. Networks constructed with traditional ego-centered and
sampling techniques are too biased to compute bridging
forces. Exhaustive graphs of small social groups will not work either,
since such groups are, by definition, dominated by bounding
relations. Since the essence of bridges is to connect individuals
across distant social regions, they can only be computed in large and
complete social graphs. Hopeless until a few years ago, such endeavor
seems more and more reasonable as digital media spread through
society. Thanks to digital traceability it is now possible to draw
large and even huge social networks
\cite{Vespignani2009,Lazer2009,Venturini2010}.

Let's discuss now the second point, the ``theory'' needed to measure the
bridging force of different edges or nodes\footnote{In this paper, we
  will focus on defining the bridgeness of nodes, but our definition
  can straightforwardly be extended to edges, just as the betweenness
  of edges is derived from that of nodes.}. Being able to identify
bounding and bridging nodes has a clear interest for any type of
network. In social networks, bounding and bridging measures (or
''closure'' and ``brokerage'', to use Burt's terms \cite{Burt2005}) tell
us which nodes build social territories and which allow items (ideas,
pieces of information, opinions, money...) to travel through them. In
scientometrics' networks, these notions tell us which authors define
disciplines and paradigms and which breed interdisciplinarity. In
ecological networks, they identify relations, which create specific
ecological communities and the ones connecting them to larger
habitats.

In all these contexts, it is the very same question that we wish to
ask: do nodes or edges reinforce the density of a cluster of nodes
(bounding) or do they connect two separated clusters (bridging)?
Formulated in this way, the bridging/bounding question seems easy to
answer. After having identified the clusters of a network, one should
simply observe if a node connects nodes of the same cluster (bounding)
or of different clusters (bridging). However, the
intra-cluster/inter-cluster approach is both too dependent on the
method used to detect communities and flawed by its inherent circular
logic: it uses clustering to define bridging and bounding ties when it
is precisely the balance of bridges and bounds that determines
clusters. Remark that, far from being a mathematical subtlety, this
question is a key problem in social theory. Defining internal
(gemeinschaft) and external (gesellschaft) relations by presupposing
the existence and the composition of social groups is absurd as groups
are themselves defined by social relations.

In this paper, we introduce a measure of bridgeness of nodes that is
independent on the community structure and thus escapes this vicious
circle, contrary to other proposals
\cite{Nepusz2008,Cheng2010}. Moreover, since the computation of
bridgeness is straightforwardly related to that of the usual
betweenness, Brandes' algorithm \cite{Brandes2001} can be used to
compute it efficiently\footnote{We have written a plug-in for Gephi
  \cite{Bastian2009} that computes this measure on large graphs. See
  Supplementary Informations for a pseudo-algorithm for both node and edge
  bridgeness.}. To demonstrate the power of our method and identify
nodes acting as local or global bridges, we apply it on a synthetic
network and two real ones: the world airport network and a
scientometric network.

\section*{Measuring bridgeness}

\begin{figure}
    \centering\includegraphics[width=0.9\textwidth]{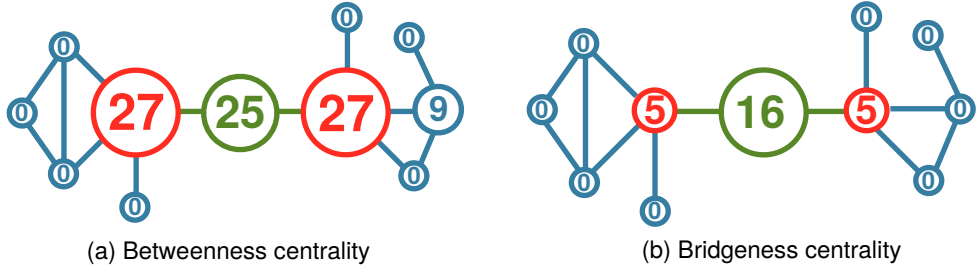}
    \caption{The figures show the betweenness (a) and bridgeness (b)
      scores for a simple graph. Betweenness does not distinguish
      centers from bridges, as it attributes a slightly higher score
      (Figure a, scores = 27) to high-degree nodes, which are local
      centers, than to the global bridge (Figure a, score = 25). In
      contrast, bridgeness rightly spots out the node (Figure b, score
      = 16) that plays the role of a global bridge.}
    \label{fig:1}
\end{figure}

Identifying important nodes in a network structure is crucial for the
understanding of the associated real-world system
\cite{Bonacich1987,Borgatti2005,Estrada2005}, for a review see
\cite{Newman2010}. The most common measure of centrality of a node for
network connections on a global scale is betweenness centrality
($BC$), which ``measures the extent to which a vertex lies on paths
between other vertices'' \cite{Freeman1977,Freeman1979}. We show in
the following that, when trying to identify specifically \emph{global}
bridges, $BC$ has some limitations as it assigns the same importance 
to paths between the immediate neighbours of a node as to paths between 
further nodes in the network. In other words BC is built to capture 
the overall centrality of a node, and is not specific enough to
distinguish between two types of centralities: local (center of a
community) and global (bridge between communities). Instead, our
measure of bridging is more specific, as it gives a higher score to
global bridges. The fact that $BC$ may attribute a higher score to
local centers than to global bridges is easy to see in a simple
network (Figure \ref{fig:1}). The logics is that a ``star'' node with
degree $k$, \emph{i.e.} a node without links between all its first
neighbors (clustering coefficient 0) receives automatically a $BC$ =
$k(k-1)/2$ arising from paths of length 2 connecting the node's first
neighbors and crossing the central node. More generally, if there
exist nodes with high degree but connected only locally (to nodes of
the same community), their betweenness may be of the order of that
measured for more globally connected nodes. Consistent with this
observation, it is well-known that for many networks, $BC$ is highly
correlated with degree \cite{Nakao1990, Goh2003, Newman2005}. A recent
scientometrics study tried to use betweenness centrality as ``an
indicator of the interdisciplinarity of journals'' but noted that this
idea only worked ``in local citation environments and after
normalization because otherwise the influence of degree centrality
dominated the betweenness centrality measure \cite{Leydesdorff2007}.

To avoid this problem and specifically spot out global centers, we
decompose $BC$ into a local and a global term, the latter being called
'bridgeness' centrality. Since we want to distinguish global bridges
from local ones, the simplest approach is to discard shortest paths,
which either start or end at a node's first neighbors from the
summation to compute $BC$ (Eq.~\ref{eq:bc}). This completely removes
the paths that connect two non connected neighbors for 'star nodes'
(see Figure \ref{fig:1}) and greatly diminishes the effect of high
degrees, while keeping those paths that connect more distant regions
of the network.

More formally in a graph $\mathcal G=(V,E)$, where $V$ assigns the set
of nodes and $E$ the set of links the definition of the betweenness
centrality for a node $j\in V$ stands as:

\begin{equation}
BC(j) =Bri(j) + local(j),
  \label{eq:bc}
\end{equation}
where
\begin{align}
\begin{split}
BC(j) & =\sum_{i\neq j\neq k}\frac{\sigma_{ik}(j)}{\sigma_{ik}} \\
\
Bri(j) & =\sum_{\substack{i \not \in N_G(j)  \wedge  k  \not \in N_G(j)}}\frac{\sigma_{ik}(j)}{\sigma_{ik}} \\
\
local(j) & =\sum_{\substack{i \in N_G(j) \vee k \in N_G(j)}}\frac{\sigma_{ik}(j)}{\sigma_{ik}}.
  \label{eq:bcdecomp}
\end{split}
\end{align}
Here the summation runs over any distinct node pairs $i$ and $k$;
$\sigma_{ik}$ represents the number of shortest paths between $i$ and
$k$; while $\sigma_{ik}(j)$ is the number of such shortest paths
running through $j$. Decomposing $BC$ into two parts (right hand side)
the first term defines actually the global term, \textit{bridgeness
centrality}, where we consider shortest paths between nodes not in the
neighbourhood of $j$ ($N_G(j)$), while the second \textit{local term} considers
the shortest paths starting or ending in the neighbourhood of
$j$. This definition also demonstrates that the bridgeness centrality
value of a node $j$ is always smaller or equal to the corresponding
$BC$ value and they only differ by the local contribution of the first
neighbours. Fig.~\ref{fig:1} illustrates the ability of bridgeness to
specifically highlight nodes that connect different regions of a
graph. Here the $BC$ (Fig.~\ref{fig:1}a) and bridgeness centrality
values (Fig.~\ref{fig:1}b) calculated for nodes of the same network
demonstrate that bridgeness centrality gives the highest score to the
node which is central globally (green), while $BC$ does not
distinguish among local or global centers, and actually assigns the
highest score to nodes with high degrees (red).

In the following, to further explore the differences between these
measures we define an independent reference measure of bridgeness
using a known partitioning of the network. This measure provides us an
independent ranking of the bridging power of nodes, that we correlate
with the corresponding rankings using the $BC$ and bridgeness
values. In addition we demonstrate via three example networks that
bridgeness centrality is always more specific than $BC$ to identify
global bridges.

\section*{Computing global bridges from a community structure}

To identify the global bridges independently from their score in $BC$
or bridgeness, we use a simple indicator inspired by the well-known
Rao-Stirling index \cite{Rao1982, Stirling2007, Rafols2014,
  Jensen2013}, as this indicator is known to quantify the ability of
nodes to connect different communities. Moreover, it includes the
notion of ``distance'', which is important for distinguishing local and
global connections. However, we note that this index needs as input a
prior categorization of the nodes into distinct communities. Our
global indicator $G$ in Eq.\ref{eq:stirling} for node $i$ is defined
as:
\begin{equation}
  G(i)=\sum_{J \in communities} l_{IJ}\delta_{i,J}
  \label{eq:stirling}
\end{equation}
where the sum runs over communities $J$ (different from the community of
node $i$, taken as $I$), $\delta_{i,J}$ being $1$ if there is a link
between node $i$ and community $J$ and $0$ otherwise. Finally,
$l_{IJ}$ corresponds to the 'distance' between communities $I$ and
$J$, as measured by the inverse of the number of links between them:
the more links connect two communities, the closer they are. Nodes
that are only linked to nodes of their own community have $G = 0$,
while nodes that connect two (or more) communities have a strictly
positive indicator. Those nodes that bridge distant communities, for
example those that are the only link between two communities, have
high $G$ values.

As a next step we use this reference measure (\emph{i.e.} the global
indicator) to rank nodes and compare it to the rankings obtained by
the two tentative characteristics of bridging ($BC$ and bridgeness) in
three large networks.

\section*{Synthetic network: unbiased LFR}

We start with a synthetic network obtained by a method similar to that
of Lancichinetti et al \cite{Lancichinetti2008}. This method leads to
the so-called 'LFR' networks with a clear community structure, which
allows to easily identify bridges between communities. We have only
modified the algorithm to obtain bridges without the degree bias which
arises from the original method. Indeed, LFR first creates unconnected
communities and then chooses randomly internal links that are
reconnected outside the community. This leads to bridges, \emph{i.e.}
nodes connected to multiple communities, which have a degree
distribution biased towards high degrees. In our method, we avoid this
bias by randomly choosing nodes, and then one of their internal links,
which we reconnect outside its community as in LFR. As reference, we
use the global indicator defined above. As explained, this indicator
depends on the community structure, which is not too problematic here
since, by construction, communities are clearly defined in this
synthetic network.

\begin{figure}
    \centering\includegraphics[width=0.5\textwidth]{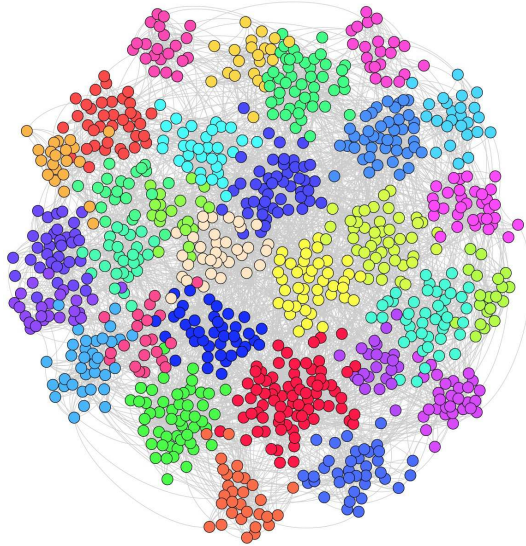}
    \caption{Artificial network with a clear community structure using
      Lancichinetti et al \cite{Lancichinetti2008} method. For
      clarity, we show here a smaller network containing 1000 nodes,
      30 communities, 7539 links ($20\%$ inter-and $80\%$
      intra-community links). Each color corresponds to a community as
      detected by modularity optimization \cite{Newman2010,Blondel2008}.}
    \label{fig:2}
\end{figure}

\begin{figure}
    \centering\includegraphics[width=0.9\textwidth]{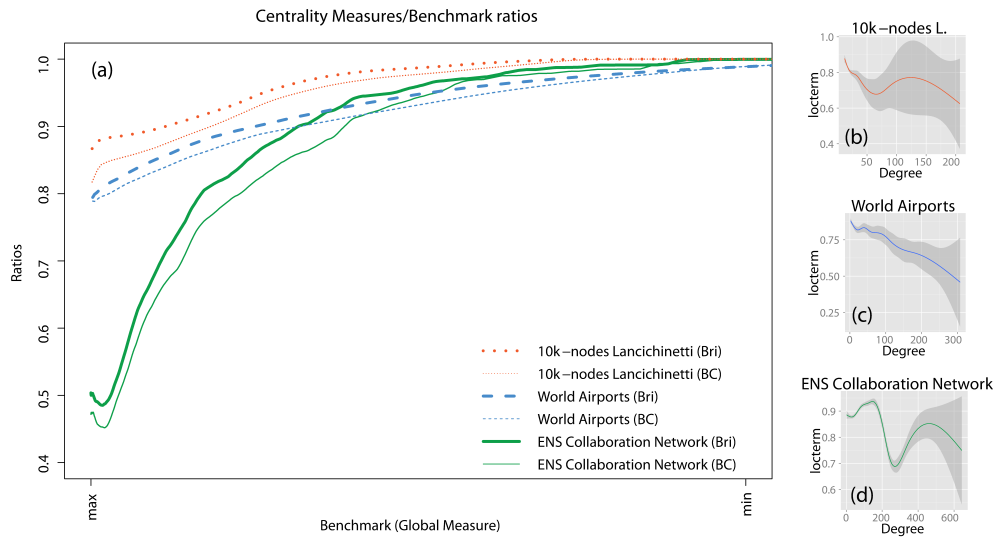}
    \caption{(a) Ability of BC or bridgeness to reproduce the ranking of
      bridging nodes, taking as reference the global indicator (Eq
      2). For each of the three networks, we first compute the
      cumulative sums for the global measure G, according to three
      sorting options: the G measure itself and the two centrality
      metrics, namely BC and bridgeness. By construction, sorting by G
      leads to the highest possible sum, since we rank the nodes
      starting by the highest G score and ending by the lowest. Then
      we test the ability of BC or bridgeness to reproduce the ranking
      of bridging nodes by computing the respective ratios of their
      cumulative sum, ranking by the respective metric (BC or Bri), to
      the cumulative obtained by the G ranking. A perfect match would
      therefore lead to a ratio equal to 1. Since we observe that the
      ratio for bridgeness is higher than for BC, this means that
      ordering nodes by their decreasing bridgeness leads to a better
      ranking of the 'global' scores as measured by G. To smooth the
      curves, we have averaged over 200 points. Curves corresponding to different networks are colorised as LFR (red), Airports (blue), ENS (green). (b, c, d): average relative local terms as function of node degree for the three investigated networks (for definition see text).}
    \label{fig:3}
  \end{figure}

Fig.~\ref{fig:3}a shows that bridgeness provides a ranking that is
closer to that of the global indicator than $BC$. Indeed, we observe
that the ratio for bridgeness is higher than for $BC$. This means that
ordering nodes by their decreasing bridgeness leads to a better
ranking of the 'global' scores - as measured by G - than the
corresponding ordering by their decreasing $BC$ values. As shown in
the simpler example of a 1000-node network (demonstrated in Fig.~\ref{fig:2}), $BC$ fails because it
ranks too high some nodes that have no external connection but have a
high degree. A detailed analysis of the nodes of a cluster is given in
Supplementaty Informations.

In addition we directly measured $\langle locterm \rangle_{i}(k)=\langle (BC(i,k) - Bri(i,k))/BC(i,k)\rangle_i$, the average relative contribution of the local term in $BC$ for nodes of the same degree (see Fig.~\ref{fig:3}b). We observe a negative correlation, which means that the local term is dominating for low degree nodes, while high degree nodes have higher bridgeness value as they have a higher chance to connect to different communities.

\section*{Real network 1: airport's network}

Proving the adequacy of bridgeness to spot out global bridges on real
networks is more difficult, because generally communities are not
unambiguously defined, therefore neither are global bridges. Then, it
is difficult to show conclusively that bridgeness is able to
specifically spot these nodes.  To answer this challenge, our strategy
is the following:

(i) We use flight itinerary data providing origin destination pairs
between commercial airports in the world (International Air Transport
Association). The network collects 47,161 transportation connections
between 7,733 airports. Each airport is assigned to its country.

(ii) We consider each country to be a distinct 'community' and compute
a global indicator based on this partitioning, as it allows for an
objective (and arguably relevant) partition, independent from any
community detection methods. Then we show that bridgeness offers a
better ranking than BC to identify airports that act as global
bridges, \emph{i.e.} that connect countries internationally.

  \begin{figure}
    \centering\includegraphics[width=0.65\textwidth]{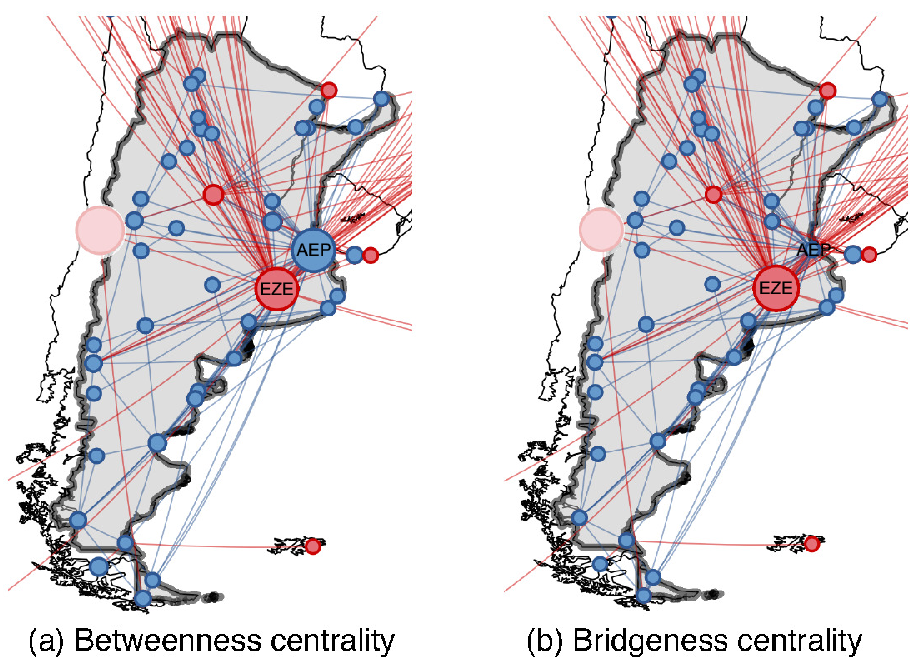}
    \caption{Example of the two largest Argentinean airports, Ezeiza
    (EZE) and Aeroparque (AEP). Both have a similar degree (54 and 45
    respectively), but while the first connects Argentina to the rest
    of the world ($85\%$ of international connections, average
    distance 2.848 miles, G=2327.2), Aeroparque is only a local center
    ($18\%$ of international connections, average distance 570 miles,
    G=9.0). However, as in the simple graph (Figure 1), BC gives the
    same score to both ($BC_{EZE}$=79,000 and $BC_{AEP}$ = 82,000),
    while bridgeness clearly distinguishes the local center and the
    bridge to the rest of the world, by attributing to the global
    bridge a score 250 times higher ($Bri_{EZE}$=46,000 and
    $Bri_{AEP}$ = 174). Red nodes represent international airports
      while blue nodes are domestic.}
    \label{fig:4}
  \end{figure}

As an example, in Fig.~\ref{fig:4} we show the two largest airports of
Argentina, Ezeiza (EZE) and Aeroparque (AEP). Both have a similar
degree (54 and 45 respectively), but while the first connects
Argentina to the rest of the world, Aeroparque mostly handles domestic
flights, thus functioning as a local center. This is confirmed by the
respective G values: 2327.2 (EZE) and 9.0 (AEP). However, just like in
our simple example in Fig.~\ref{fig:1}, $BC$ gives the same score to
both, while bridgeness clearly distinguishes between the local
domestic center and the global international bridge by attributing to
the global bridge a score 250 times higher (see
Fig.~\ref{fig:4}). This can partly be explained by the fact that AEP
is a 'star' node (low clustering coefficient: 0.072), connected to 12
very small airports, for which it is the only link to the whole
network. All the paths starting from those small airports are
cancelled in the computation of the bridgeness (they belong to the
\emph{'local'} term in Eq.\ref{eq:bc}), while $BC$ counts them equally
as any other path.

More generally, Figure 3 shows that, as for the Airport network,
bridgeness provides again a ranking that is closer to that of the global
indicator. Indeed, ordering nodes by their decreasing bridgeness leads
to a ranking that is closer to the ranking obtained by the global
score than the ranking by decreasing $BC$. In addition we found again negative correlations between the average relative local term and node degrees (see Fig.~\ref{fig:3}c), assigning similar roles for low and high degree nodes as in case of the synthetic network.

\section*{Real network 2: scientometric network of ENS Lyon}

\begin{figure}
    \centering\includegraphics[width=0.9\textwidth]{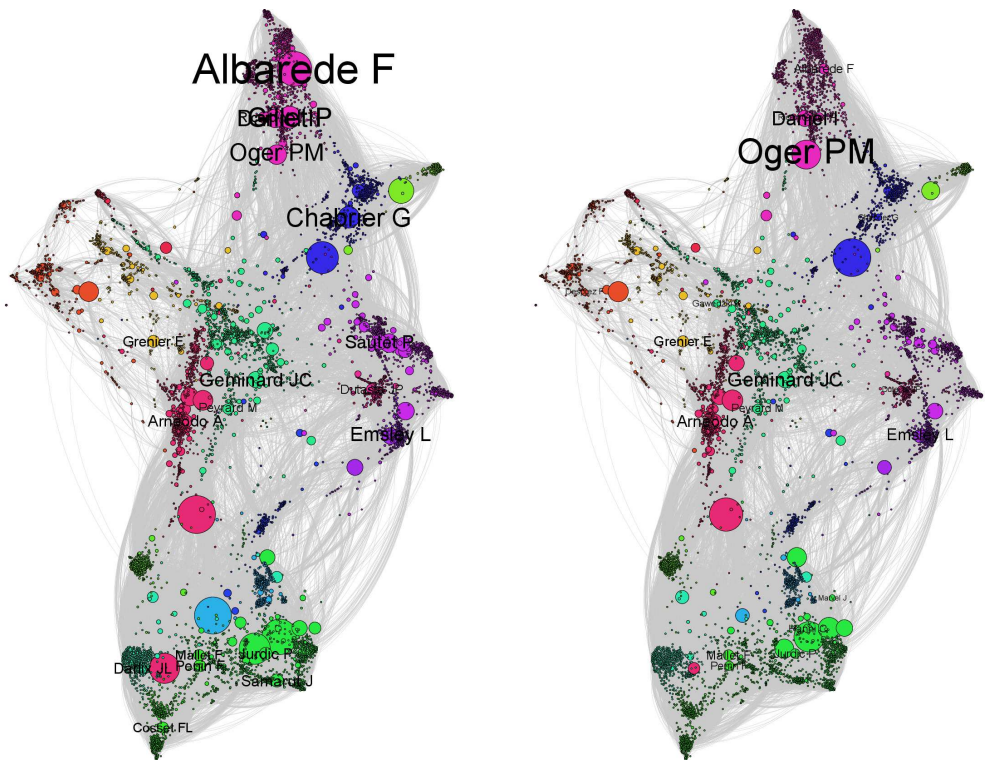}
    \caption{Co-citation and co-author network of articles published
      by scientists at ENS de Lyon. Nodes represent the authors or
      references appearing in the articles, while links represent
      co-appearances of these features in the same article. The color
      of the nodes corresponds to the modularity partition and their
      size is proportional to their $BC$ (left) or to their bridgeness (right),
      which clearly leads to different rankings (references cited are
      used in the computations of the centrality measures but appear
      as dots to simplify the picture). We only keep nodes that appear
      on at least four articles and links that correspond to at least
      2 co-appearances in the same paper. After applying these
      thresholds, the 8000 articles lead to 8883 nodes (author or
      references cited in the 8000 articles) and 347,644 links. The
      average degree is 78, the density 0.009 and the average
      clustering coefficient is 0.633. Special care was paid to avoid
      artifacts due to homonyms. Weights are attributed to the links
      depending on the frequency of co-appearances (cosine distance,
      see \cite{Grauwin2011}.}
    \label{fig:5}
  \end{figure}

The second example of a real network is a scientometric graph of a
scientific institution \cite{Grauwin2011}, the ``Ecole normale
sup\'erieure de Lyon'' (ENS, see Figure \ref{fig:5}). This
networks adds authors to the usual co-citation network, as we want to
understand which authors connect different sub-fields and act as
global, interdisciplinary bridges. To identify the different
communities, we rely on modularity optimization \cite{Blondel2008}, which leads to a
relevant community partition because scientific networks are highly
structured by disciplinary boundaries. This is confirmed by the high
value of modularity generated by this partition (0.89). In Figure
\ref{fig:5}, the authors of different communities are shown
with different colors, and their size corresponds to their betweenness (left) or
bridgeness (right) centrality, which clearly leads to highlight different
authors as the main global bridges, which connect different
subfields. We compute the Stirling indicator (Eq.\ref{eq:bc}) based on
the modularity structure to identify the global bridges. As for the
previous networks, Fig.~3 shows that bridgeness ranks the nodes in a
closer way than $BC$ to the ranking provided by the global measure
based on community partition. On the other hand the corresponding $\langle locterm \rangle (k)$ function (see Fig.~\ref{fig:3}d) suggests a slightly different picture in this case. Here nodes with large but moderate degrees (smaller than $\sim 200$) have high local terms suggesting that they act as local centres, while nodes with higher degrees have somewhat smaller local terms assigning their role to act as global bridges.

\section*{Discussion}

In this paper we introduced a measure to identify nodes acting as global bridges in complex network structures. Our proposed methodology is based on the decomposition of $BC$ into a local and global term, where the local term considers shortest paths that start or end at one of the node's neighbors, while the global term, what we call bridgeness, is more specific to identify nodes which are globally central. We have shown, on both synthetic and real networks, that the proposed bridgeness measure improves the capacity to specifically find out global bridges as it is able to distinguish them from local centers. One crucial advantage of our measure of
bridgeness over former propositions is that it is independent of the
definition of communities.

However, the advantage in using bridgeness depends the precise
topology of the network, and mainly on the degree distribution of
bridges as compared to that of all the nodes in the network. When
bridges are high-degree nodes, $BC$ and bridgeness give an equally good approximation, since
high-degree bias do not play an important role in this case. Instead, when some bridges have low
degrees, while some high-degree nodes act like local centers of their own
community, bridgeness is more effective to identify bridges as $BC$ gives
equally high rank to nodes with high degree, even if they are not
connected to nodes outside of their community. We demonstrated that bridgeness is systematically more
specific to spot out global bridges in all the networks we have
studied here. Although the improvement was small on average, typically $5$ to
$10\%$, even a small amelioration of a widely used measure is in itself an interesting result.

We should also note that, except on simple graphs, comparing these two
measures is difficult since there is no clear way to identify,
independently, the 'real' global bridges. We have used community
structure when communities seem clear-cut, but then we fall into the
circularity problems stressed in the introduction. Using metadata on
the nodes (\emph{i.e.}~countries for the airports) may solve this problem
but raises others, as metadata do not necessarily correspond to
structures obtained from the topology of the network, as shown
recently on a variety of networks \cite{Hric2014}. Another possible extension would be to identify overlapping communities to identify independently global bridges, as nodes involved in multiple communities, and correlate them with the actual measure, which provides a direction for future studies. However, in any case identifying global bridges remains a difficult problem as it is tightly linked to
another difficult problem, that of community detection. Decomposing
$BC$ into a local and a global term helps to improve the solution,
but many questions remain still open for further inquiry.

\nocite{*}

\bibliographystyle{comnet}
\bibliography{biblio} 

\pagebreak

\section*{\Large{Supplementary Informations}}

\section*{S1. Modified Brandes algorithm}

\setcounter{table}{0} \setcounter{figure}{0}
\renewcommand{\thetable}{S\arabic{table}}%
\renewcommand{\thefigure}{S\arabic{figure}}%

\bigskip

Bridgeness algorithm, inspired by Brandes' ``faster algorithm''
\cite{Brandes2001}

\bigskip

SP[s,t] $\leftarrow $precompute all shortest distances
matrix/dictionary

CB[v] $\leftarrow $ 0, v ${\in}$ V ;

\textbf{for} s ${\in}$ V \textbf{do}

\ \ S $\leftarrow $ empty stack;

\ \ P[w] $\leftarrow $ empty list, w ${\in}$ V ;

\ \ $\sigma $[t] $\leftarrow $ 0, t ${\in}$ V ; $\sigma $[s]
$\leftarrow $ 1;

\ \ d[t] $\leftarrow $ $-$1, t ${\in}$ V ; d[s] $\leftarrow $ 0;

\ \ Q $\leftarrow $ empty queue;

\ \ enqueue s $\rightarrow $ Q;

\ \ \textbf{while} Q not empty \textbf{do}

\ \ \ \ dequeue v $\leftarrow $ Q;

\ \ \ \ push v $\rightarrow $ S;

\ \ \ \ \textbf{foreach} neighbor w of v \textbf{do}

\ \ \ \ \ \ // w found for the first time?

\ \ \ \ \ \ \textbf{if} d[w] {\textless} 0 \textbf{then}

\ \ \ \ \ \ \ \ enqueue w $\rightarrow $ Q;

\ \ \ \ \ \ \ \ d[w] $\leftarrow $ d[v] + 1;

\ \ \ \ \ \ \textbf{end}

\ \ \ \ \ \ // shortest path to w via v?

\ \ \ \ \ \ \textbf{if} d[w] = d[v] + 1 \textbf{then}

\ \ \ \ \ \ \ \ $\sigma $[w] $\leftarrow $ $\sigma $[w] + $\sigma
$[v];

\ \ \ \ \ \ \ \ append v $\rightarrow $ P[w];

\ \ \ \ \ \ \textbf{end}

\ \ \ \ \textbf{end}

\ \ \textbf{end}

\ \ $\delta $[v] $\leftarrow $ 0, v ${\in}$ V ;

\ \ // S returns vertices in order of non-increasing distance from s

\ \ \textbf{while} S not empty \textbf{do}

\ \ \ \ pop w $\leftarrow $ S;

\ \ \ \ \textbf{for} v ${\in}$ P[w] \textbf{do} $\delta $[v]
$\leftarrow $ $\delta $[v] + $\sigma $[v]/$\sigma $[w]
{\textperiodcentered} (1 + $\delta $[w]);

\ \ \ \ \textbf{if} SP[w,s]{\textgreater}1 \textbf{then} CB[w]
$\leftarrow $ CB[w] + $\delta $[w];

\ \ \textbf{end}

\textbf{end}

\section*{S2. Case study on a synthetic network community}

The specificity of bridgeness and the influence of the degree, which
prevents BC from identifying correctly the most important bridges, can
be exemplified by examining the scores of nodes in cluster 5 of the
synthetic network. This cluster is linked to cluster 13 by 5
connections (through nodes 248, 861, 471, 576 and 758) and to cluster
1 by a single connection (through node 232). BC gives roughly the same
score to nodes 232 and 248, while bridgeness attributes a score almost
4 times higher to node 232, correctly pointing out the importance of
this single bridge between clusters 5 and 1. This is because BC is
confused by the high degree of node 248 (41) as compared to node 232
low degree (20). Therefore, by counting all the shortest paths, BC
attributes too high a bridging score to node 248. Second problem with
BC, it gives a high score to nodes that are not connected to other
communities, merely because they are local centers, \emph{i.e.} they
have a high degree. For example, node 515 obtains a higher BC score
than node 758 (Table \ref{tab:s1}), even if node 515 has no connection
to other communities (but degree 49), contrary to node 758 (connected
to cluster 5, but degree 23). Bridgeness never ranks higher local
centers than global bridges: here, it correctly assigns a 5 times
higher score to node 758 than to node 515.

\begin{center}
  \begin{figure}
    \includegraphics[width=0.65\textwidth]{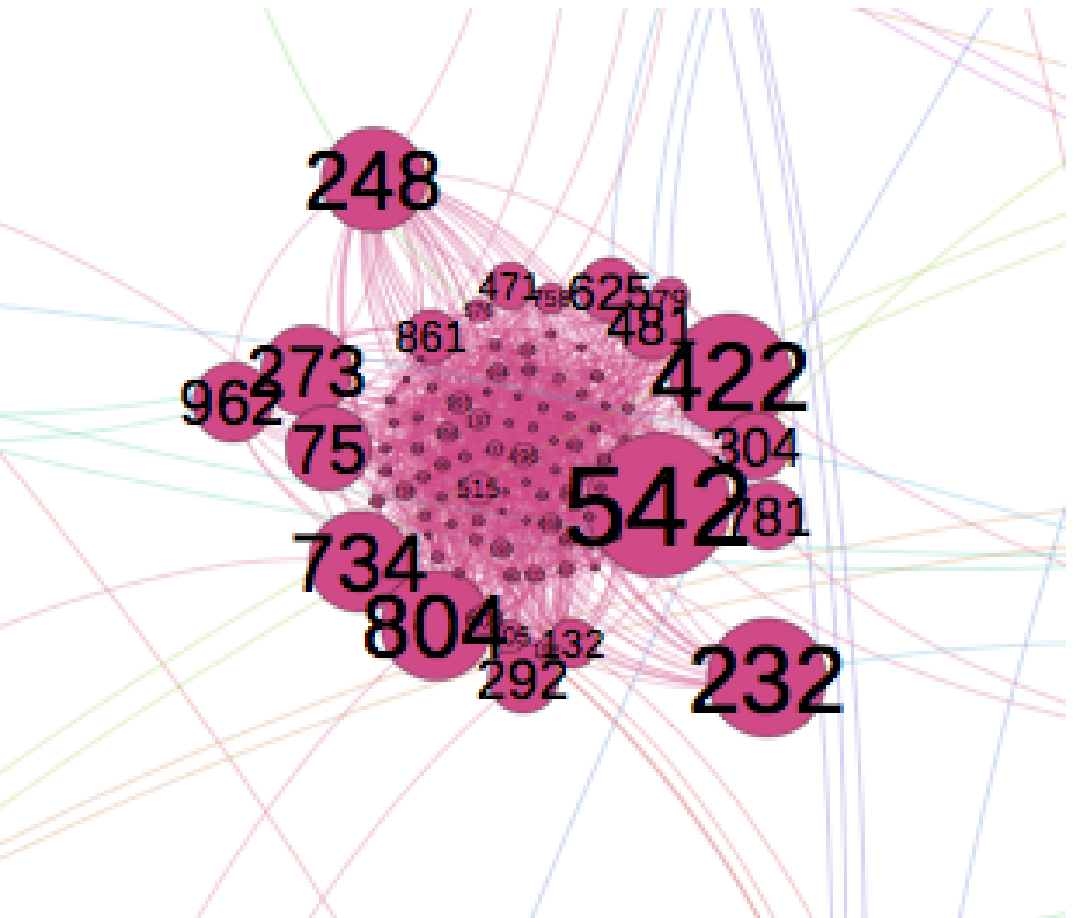}
    \caption{Zoom on cluster 5 of the synthetic network. The numbers
      show node's labels, while the size of the nodes is proportional
      to their BC score.}
    \label{fig:s1}
  \end{figure}
\end{center}

\begin{table}[t!]
  \caption{Nodes in community 5 of the synthetic network, ranked by decreasing BC (see text)}
  \label{tab:s1}
  \setlength{\tabcolsep}{5pt}
  \begin{center}
    \begin{tabular}{l || r | r | r | r | r}
      Id & Stirling & Modularity Class & Betweenness & Bridgeness & Degree \\
      \hline
      542 & 0.0222 & 5 & 9173.71 & 2644.62 & 44 \\
      422 & 0.0278 & 5 & 7714.27 & 3855.62 & 35 \\
      232 & 0.0950 & 5 & 7551.22 & 5846.86 & 20 \\
      804 & 0.0285 & 5 & 6995.63 & 2824.64 & 34 \\
      248 & 0.0082 & 5 & 6588.65 & 1624.30 & 48 \\
      734 & 0.0907 & 5 & 6410.31 & 4373.72 & 21 \\
      273 & 0.0322 & 5 & 5698.28 & 2631.59 & 30 \\
      75 & 0.0868 & 5 & 5349.47 & 3558.31 & 22 \\
      962 & 0.0399 & 5 & 4989.66 & 2951.45 & 24 \\
      292 & 0.0399 & 5 & 4377.77 & 1939.06 & 24 \\
      481 & 0.0256 & 5 & 4305.68 & 1796.92 & 25 \\
      781 & 0.0475 & 5 & 4257.93 & 2200.21 & 20 \\
      304 & 0.0434 & 5 & 4221.64 & 2467.65 & 22 \\
      625 & 0.0202 & 5 & 3964.21 & 1314.62 & 32 \\
      861 & 0.0108 & 5 & 3295.01 & 714.44 & 36 \\
      132 & 0.0200 & 5 & 2985.45 & 1157.49 & 24 \\
      471 & 0.0154 & 5 & 2865.07 & 1296.38 & 25 \\
      79 & 0.0302 & 5 & 2256.02 & 1004.28 & 21 \\
      205 & 0.0208 & 5 & 1921.65 & 788.51 & 23 \\
      515 & 0.0000 & 5 & 1884.07 & 86.45 & 49 \\
      758 & 0.0166 & 5 & 1791.80 & 435.66 & 23 \\
      608 & 0.0200 & 5 & 1777.54 & 522.75 & 24 \\

    \end{tabular} 
  \end{center}

\end{table}

\end{document}